# Energy spectrum for a short-range 1/r singular potential with a non-orbital barrier using the asymptotic iteration method

A. J. Sous[1] and A. D. Alhaidari[2]

[1]*Al-Quds Open University, Tulkarm, Palestine*
[2]*Saudi Center for Theoretical Physics, P. O. Box 32741, Jeddah 21438, Saudi Arabia*

**Abstract.** Using the asymptotic iteration method, we obtain the *S*-wave solution for a short-range three-parameter central potential with 1/r singularity and with a non-orbital barrier. To the best of our knowledge, this is the first attempt at calculating the energy spectrum for this potential, which was introduced by H. Bahlouli and A. D. Alhaidari and for which they obtained the "potential parameter spectrum". Our results are also independently verified using a direct method of diagonalizing the Hamiltonian matrix in the J-matrix basis.

## 1. Introduction

In a previous article [1], we used the Asymptotic Iteration Method (AIM) to find the energy spectrum for the hyperbolic single wave potential, which was introduced by Bahlouli and Alhaidari [2,3]. In the present work, we apply the same technique in [1] to the short-range three-parameter central potential, which was introduced by the same authors [3,4] as

$$V(r) = V_0 \frac{e^{-\lambda r} - \gamma}{e^{\lambda r} - 1}, \tag{1}$$

where $V_0$ is the potential strength and the range parameter $\lambda$ is positive with an inverse length units. The dimensionless parameter $\gamma$ is in the open range $0 < \gamma < 1$. This potential is short-range with $1/r$ singularity at the origin. It is also interesting to note that, at short distance and with $0 < \gamma < 1$, there is a clear resemblance of this potential with $V_0 > 0$ to the attractive Coulomb potential [5] with non-zero angular momentum (see Figure 1). However, the potential valley here is not due to the centrifugal force attributed to the angular momentum. Moreover, it does not have the long-range behavior of the Coulomb potential [3,4]. In [4], Alhaidari argued that in contrast to the Coulomb potential the number of bound states for this potential is finite and that it could be used as a more appropriate model for the description of an electron interacting with an extended molecule whose electron cloud is congregated near the center of the molecule. The Authors of [2-4] found the "potential parameter spectrum" (PPS) for the hyperbolic single wave potential and for potential (1). The concept of a PPS was introduced for the first time in the solution of the wave equation in [6] where for a given energy the problem becomes exactly solvable for a discrete set (finite or infinite) of values of the potential parameters. If the map that associates the parameter spectrum with the energy is invertible, then in principle one could obtain the energy spectrum for a given choice of potential parameters [6].

The calculation of energy eigenvalues is one of the basic problems of elementary quantum mechanics. Many techniques have been suggested to obtain the energy spectrum associated with a given potential. We will not try to give an overview here but one may consult [7-15] and references therein for the description of the AIM, which has been used successfully in the past. In this work, we will be using this method to solve the time-independent S-wave (zero angular momentum) Schrödinger equation with potential (1) in order to find the energy spectrum.

The paper has the following structures. In Section 2, we briefly present an overview of the AIM and apply it to obtain the energy spectrum for potential (1). In Sec. 3, we find the eigen-energies where we make a comparison between the results obtained by the AIM and those obtained by the potential parameter spectrum method (PPSM). Additionally, we confirm our results independently by diagonalizing the Hamiltonian matrix, which is calculated in the J-matrix basis [16]

## 2. Basics of the AIM and its Application to Potential (1)

In this section, we present the basic concepts of the AIM. For more details, we refer the reader to [7-15]. The method could be used to solve a second-order homogeneous linear differential equations of the form

$$y''(x) = k_0(x)y'(x) + z_0(x)y(x)\,, \tag{2}$$

where $k_0(x)$ and $z_0(x)$ are continuously differentiable functions over the defined interval of the coordinate $x$. According to the asymptotic aspect of the method and for sufficiently large $n$ we have

$$\frac{z_n(x)}{k_n(x)} = \frac{z_{n-1}(x)}{k_{n-1}(x)} = \phi(x)\,, \tag{3}$$

where the function $\phi(x)$ is independent of $n$ and where

$$\begin{aligned} k_n(x) &= k'_{n-1}(x) + z_{n-1}(x) + k_0(x)k_{n-1}(x) \\ z_n(x) &= z'_{n-1}(x) + z_0(x)k_{n-1}(x) \end{aligned} \tag{4}$$

The prime stands for the derivative with respect to $x$. The general solution of Eq. (2) is obtained as

$$y(x) = \exp\left[-\int_x \phi(x')dx'\right]\left\{C_2 + C_1\int_x \exp\left[\int_{x'}\left(k_0(x'') + 2\phi(x'')\right)dx''\right]dx'\right\}. \tag{5}$$

The energy eigenvalues, $E$, of the Schrödinger equation corresponding to Eq. (2) are obtained using the iteration terminating condition by means of Eq. (3) that reads

$$\Delta_n(x) = k_{n-1}(x)z_n(x) - z_{n-1}(x)k_n(x)\,, \tag{6}$$

If the problem is analytically solvable (exact solution) then $\Delta_n(x) = 0$. In this case, the AIM gives the eigenvalues and eigenfunctions in explicit algebraic form. However, in this case there are a limited number of potentials and the condition $\Delta_n(x) = 0$ is satisfied at every point $x$ in the defined interval. That is, $\Delta_n(x)$ is independent of $x$ and the eigenvalues are determined from the $n$ roots of the condition $\Delta_n(x) = 0$. In case of approximation, $\Delta_n(x)$ depends on both $x$ and $E$. So, we have to determine a proper initial point $x = x_0$ value for solving $\Delta_n(x_0) = 0$ with respect to $E$. Ideally, the solution should be independent of the choice of $x_0$. However, an acceptable range of stability of the calculated energy eigenvalue may be found near the minimum value of the potential or the maximum value of the asymptotic wavefunction [14-15]. Next, we apply the method to the problem with the potential function (1).

Inserting potential (1) in the time-independent radial Schrödinger's equation results in the following second order linear differential equation

$$\left[-\frac{1}{2}\frac{d^2}{dr^2}+\frac{\ell(\ell+1)}{2r^2}+V_0\frac{e^{-\lambda r}-\gamma}{e^{\lambda r}-1}-E\right]\psi(r)=0\,, \tag{7}$$

where $\ell$ is the angular momentum quantum number and we have adopted the atomic units $\hbar=m=1$. Defining the new variable $x=1-2e^{-\lambda r}$, whose range is between $-1$ and $+1$, transforms this equation for $S$-wave ($\ell=0$) into the desired equation (2) where we can apply the AIM with

$$k_0(x)=\frac{1}{1-x}\,, \tag{8}$$

$$z_0(x)=\frac{2V_0}{\lambda^2}\left[\frac{1/2}{1+x}-\frac{\gamma}{1-x^2}-\frac{E/V_0}{(1-x)^2}\right]. \tag{9}$$

Using these seed functions together with the quantization condition (6) in the iteration, we can calculate the energy eigenvalues. Since the problem is not exactly solvable, we have to select a proper $x_0$. In this work, we observed that the best initial point is $x_0=0$, which corresponds to the middle of the $x$ interval. Therefore, at the end of the iterations when a stable result is reached, the energy spectrum is obtained by substituting $x=0$.

## 3. Results and Discussion

The calculation of the energy eigenvalues for potential (1) is not as easy as that for the hyperbolic single wave potential, which we have studied in [1]. This is because the present potential has $1/r$ singularity at the origin. In this work, we take the value of the parameter $\gamma$ between zero and one. If $\gamma$ is less than zero or larger than one then the potential is not too interesting since it will not have the potential valley or potential hill (see Fig. 1) and will only be similar to the well-known Hulthen or Yukawa potentials [17]. The necessary but not sufficient conditions for the existence of bound states for this potential are as follows:

1. If $\gamma$ is between zero and one, then any value of $V_0$ positive or negative could in principle support bound states.
2. If $\gamma$ is greater than one or less than zero, then the sign of $V_0$ must be the same as that of $\gamma$.

In Table 1, we compare the energy spectrum obtained using the AIM outlined above and those obtained by the PPSM for $V_0=5$, $\lambda=0.2$ and for several values of $\gamma$ between zero and one. The Table shows good agreement between the two results. However, the PPSM results are in better agreement with the direct method of diagonalization of the Hamiltonian matrix in the J-matrix basis (HDM). In the Appendix, we show how this diagonalization procedure is carried out. Table 2 shows also another good agreement between the results obtained by the two methods but now $\gamma$ is fixed while $V_0$ is varied.

## Appendix: Diagonalization of the Hamiltonian matrix in the J-matrix basis

The Hamiltonian of the problem is $H = T + V(r)$, where the kinetic energy operator is $T = -\frac{1}{2}\frac{d^2}{dr^2} + \frac{\ell(\ell+1)}{2r^2}$ and $V(r)$ is the potential in (1). Diagonalization of the matrix representation of this Hamiltonian in a given basis is difficult for two reasons. First, due to the singular behavior at the origin, which goes like $1/r^2$ for $T$ and like $1/r$ for $V$. Second, the long-range behavior of $T$, which dictates that we need to deal with infinite dimensional matrices. On the other hand, if we manage to transfer the $1/r$ singularity from $V$ and add it to $T$ by writing $V(r) = \frac{Z}{r} + U(r)$, where $\frac{Z}{r} = \lim_{r\to 0} V(r)$, then $Z = \frac{V_0}{\lambda}(1-\gamma)$ and we can write $H = H_0 + U(r)$ where the reference Hamiltonian $H_0 = T + \frac{Z}{r}$. Consequently, the new potential $U(r)$ is regular everywhere (see Fig. 2) and thus it could be approximated very well by its matrix elements in a finite subset of a square integrable basis. Then, what remains is to have a full account for the matrix representation of the reference Hamiltonian $H_0$. However, $H_0$ is just the Coulomb Hamiltonian, which is known to have an exact tridiagonal matrix representation in the square integrable Laguerre basis with the following elements [16]

$$\chi_n(r) = a_n (\mu r)^{\ell+1} e^{-\mu r/2} L_n^{2\ell+1}(\mu r), \tag{A1}$$

where $L_n^{\nu}(z)$ is the associated Laguerre polynomial of order $n$ in $z$ and $\mu$ is a length scale parameter. The normalization constant is chosen as $a_n = \sqrt{\Gamma(n+1)/\Gamma(n+2\ell+2)}$. This gives an infinite tridiagonal symmetric matrix representation for the reference Hamiltonian whose elements read as follows

$$\begin{gathered}\left(H_0\right)_{nm} = \\ \tfrac{\mu^2}{4}\left[\left(n+\ell+1+\tfrac{4Z}{\mu}\right)\delta_{n,m} + \tfrac{1}{2}\sqrt{n(n+2\ell+1)}\,\delta_{n,m+1} + \tfrac{1}{2}\sqrt{(n+1)(n+2\ell+2)}\,\delta_{n,m-1}\right]\end{gathered} \tag{A2}$$

As noted above, the potential $U(r)$ is easily accounted for by its matrix elements in a finite subset of the basis (A1). Increasing the size of this subset will improve the accuracy of the results. These matrix elements are obtained by evaluating the integral

$$\begin{aligned} U_{nm} &= \mu \int_0^\infty \chi_n(r) U(r) \chi_m(r)\, dr \\ &= a_n a_m \int_0^\infty x^{2\ell+1} e^{-x} L_n^{2\ell+1}(x) L_m^{2\ell+1}(x) \left[x U(x/\mu)\right] dx \end{aligned} \tag{A3}$$

where $x = \mu r$. The evaluation of this integral is performed using the Gauss quadrature approximation scheme [18], which gives

$$U_{nm} \cong \sum_{k=0}^{N-1} \Lambda_{nk} \Lambda_{mk} \left[\omega_k U(\omega_k/\mu)\right], \tag{A4}$$

for adequately large integer $N$. $\omega_k$ and $\{\Lambda_{nk}\}_{n=0}^{N-1}$ are the $N$ eigenvalues and corresponding normalized eigenvectors of the $N\times N$ tridiagonal basis overlap matrix whose elements are

$$\langle \chi_n | \chi_m \rangle = 2(n+\ell+1)\delta_{n,m} - \sqrt{n(n+2\ell+1)}\,\delta_{n,m+1} - \sqrt{(n+1)(n+2\ell+2)}\,\delta_{n,m-1}. \tag{A5}$$

In an ideal situation where $N$ is infinite the physical results should be independent of the choice of value of the numerical scale parameter $\mu$. Nonetheless, for finite calculation we should be able to find a range of values of $\mu$ within which the results are stable and accurate to the desired number of significant digits. Increasing the size of the representation $N$ will increase this range, which we refer to as the plateau of stability.

Thus, this parameter is analogous to the $x_0$ parameter in the AIM noted in Sec. 2 above. For a given set of physical parameters, we add (A2) to (A4) giving the matrix elements of the total Hamiltonian which could be diagonalized numerically for a proper value of the numerical scale parameter $\mu$ chosen from within the plateau of stability and for large enough matrix size *N*.

Table 1 and Table 2 show an excellent agreement of the results obtained by this diagonalization method (HDM) for $\ell = 0$ with the PPSM and shows good to fair agreement with the AIM results. The agreement is less pronounced for the highest energy states. In the HDM, we used a basis size of *N* = 100.

**Table Captions**

**Table 1**: Energy spectrum with $V_0 = 5$ and $\lambda = 0.2$ for various values of $\gamma$ obtained using the three methods mentioned in the text.

**Table 2**: Energy spectrum with $\gamma = 0.6$ and $\lambda = 0.5$ for various values of $V_0$. The results of the three independent methods are in good agreement.

**Figure Captions**

**Fig. 1**: The potential function of Eq. (1) for $\gamma = 0.3$ (solid), $\gamma = 0.5$ (dashed), $\gamma = 0.75$ (dashed-dotted), $\gamma = 0.9$ (dotted).

**Fig. 2**: The potential function *V*(*r*) in Eq. (1) before removing the singularity and then after removing it as *U*(*r*). We took $\gamma = 0.8$.

**Table 1**

| $n$ | method | $\gamma = 0.2$ | $\gamma = 0.4$ | $\gamma = 0.6$ |
|---|---|---|---|---|
| 0 | AIM<br>PPSM<br>HDM | -0.02600017100<br>-0.0260054988<br>-0.0260054988 | -0.1794066345<br>-0.1794066345<br>-0.1794066345 | -0.5368000468<br>-0.5368000468<br>-0.5368000468 |
| 1 | AIM<br>PPSM<br>HDM | -0.001138151383<br>-0.0002851232<br>-0.0002791687 | -0.07260826273<br>-0.0726083684<br>-0.0726083684 | -0.3182343338<br>-0.3182343338<br>-0.3182343338 |
| 2 | AIM<br>PPSM<br>HDM | | -0.01483611943<br>-0.0146815305<br>-0.0146815305 | -0.1627941441<br>-0.1627941432<br>-0.1627941432 |
| 3 | AIM<br>PPSM<br>HDM | | | -0.06232692757<br>-0.0623301374<br>-0.0623301374 |
| 4 | AIM<br>PPSM<br>HDM | | | -0.01131548300<br>-0.0103706012<br>-0.0103705985 |

**Table 2**

| $n$ | method | $V_0 = 20$ | $V_0 = 40$ | $V_0 = 60$ |
|---|---|---|---|---|
| 0 | AIM<br>PPSM<br>HDM | -2.017967507<br>-2.0179675071<br>-2.0179675071 | -4.417015612<br>-4.4170156123<br>-4.4170156123 | -6.886516026<br>-6.8865160257<br>-6.8865160257 |
| 1 | AIM<br>PPSM<br>HDM | -1.008842615<br>-1.0088426139<br>-1.0088426139 | -2.815063004<br>-2.8150630039<br>-2.8150630039 | -4.825915095<br>-4.8259150953<br>-4.8259150953 |
| 2 | AIM<br>PPSM<br>HDM | -0.3740938860<br>-0.3740996421<br>-0.3740996421 | -1.617682362<br>-1.6176823617<br>-1.6176823617 | -3.184708114<br>-3.1847081143<br>-3.1847081143 |
| 3 | AIM<br>PPSM<br>HDM | -0.06098871935<br>-0.0568940453<br>-0.0568940452 | -0.7773062284<br>-0.7773055229<br>-0.7773055229 | -1.920830979<br>-1.9208309792<br>-1.9208309792 |
| 4 | AIM<br>PPSM<br>HDM | | -0.2548982339<br>-0.2553793984<br>-0.2553793984 | -0.9991295725<br>-0.9991291965<br>-0.9991291965 |
| 5 | AIM<br>PPSM<br>HDM | | -0.04015268525<br>-0.0200363806<br>-0.0200357906 | -0.3895811680<br>-0.3897965892<br>-0.3897965892 |
| 6 | AIM<br>PPSM<br>HDM | | | -0.08198118960<br>-0.0672459224<br>-0.0672459104 |

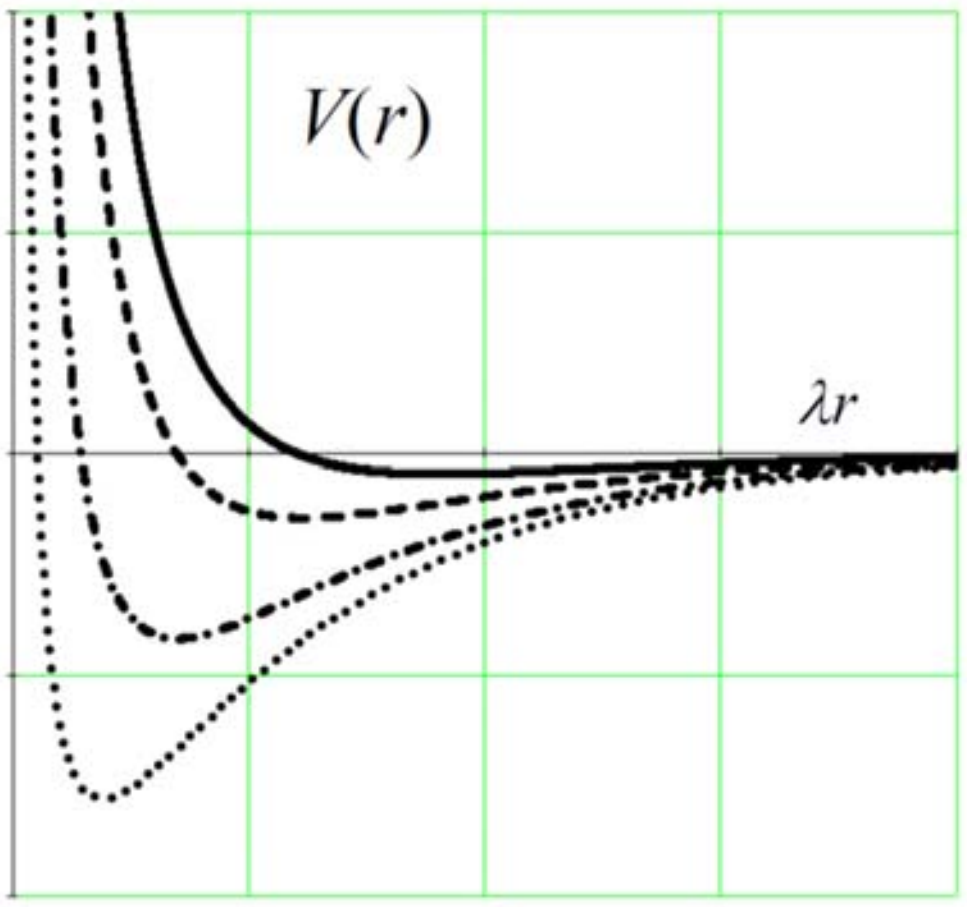

**Fig. 1**

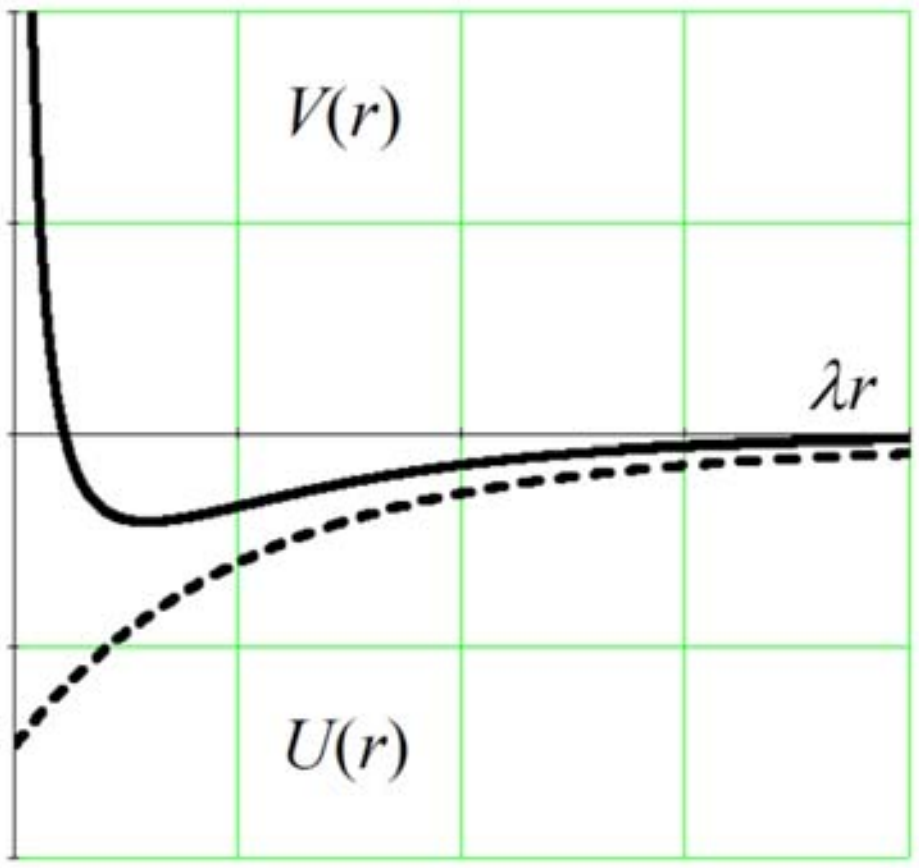

**Fig. 2**